\documentclass[fleqn,twoside]{article}
\usepackage{espcrc2,psfig}

\usepackage{graphicx}
\usepackage[figuresright]{rotating}

\newcommand{\AmS}{{\protect\the\textfont2
  A\kern-.1667em\lower.5ex\hbox{M}\kern-.125emS}}

\title{The last Gift of {\it Beppo}SAX: PDS Observations of the two Blazars 
1ES 0507-040 and PKS 1229-021}

\author{L. Chiappetti\address[CNR]{CNR/IASF, Milano (Italy)},
G. Ghisellini\address[OAB]{INAF/OAB, Milano (Italy)},
L. Maraschi\addressmark[OAB], 
E. Pian\address[OAT]{INAF/OAT, Trieste (Italy)},
F. Tavecchio\addressmark[OAB],  
A. Treves\address[INS]{Univ. Insubria, Como (Italy)}, 
A. Wolter\addressmark[OAB]}
       
\begin{document}

\begin{abstract}
Towards the end of the {\it Beppo}SAX mission, the only operated Narrow
Field Instrument was the PDS, which covers the energy range 13-300 keV.
Two blazars, 1ES 0507-040 ($z=0.304$) and PKS 1229-021 ($z= 1.045$),
suitably located in the sky for {\it Beppo}SAX pointing, were observed for
about 2 days each in Spring 2002 with the PDS and detected up to 50
keV. 1ES0507-040 had been already observed by the {\it Beppo}SAX NFIs at an
earlier epoch for a much shorter time. We have re-analysed those data as
well, and identified a possible contamination problem in the PDS
spectrum. We present our recent PDS data on both sources, combined with
the previous {\it Beppo}SAX data and with non-simultaneous observations at
other frequencies. The derived Spectral Energy Distributions allow us to
discuss the origin of the high energy component.
\vspace{1pc}
\end{abstract}

\maketitle

\section{Introduction}

The class of blazars contains the most extreme AGNs, characterized by
fast variability, high degree of polarization, bright non-thermal
continuum extending from radio up to $\gamma $-ray energies (e.g. Urry \&
Padovani 1995). The study of these sources allows us to address several
important issues related to the physics of relativistic jets. For
these reasons it is of particular importance to obtain information on the
high-energy emission of blazars. In this respect an excellent instrument
has been the Italian-Dutch satellite {\it Beppo}SAX (Boella et al. 1997).
In fact the unprecedentely wide band of {\it Beppo}SAX (from 0.1 to above
100 keV) is optimal to study the connection between the X-ray and the
$\gamma$-ray continuum and to disentangle the different emission
components. One can therefore constrain more effectively the physical 
properties of the emission regions.

In Spring 2002, the performance of {\it Beppo}SAX started to fail.
In particular, the acquisition of fields too far
from the celestial Equator became problematic. Therefore, our planned
campaign on the BL Lac Mkn 501 had to be
cancelled. Furthermore, the only NFI routinely operated (because of
battery problems) was the high-energy instrument PDS (13-300 keV). We
thus were solicited to investigate the opportunity of observing, in
replacement of Mkn 501, blazar sources with celestial locations satisfying
the limited pointing capabilities of {\it Beppo}SAX.  The only 2 sources that
could be pointed by the satellite at that epoch, and that appeared at the same
time promising in view of a PDS detection, were: the BL Lac 1ES0507-040
and the luminous Flat Spectrum Radio Quasar (FSRQ) PKS 1229-021.

In the following we present the analysis of the data and the results of
the observations (Sect 2) and the discussion (Sect 3).

\section{Data analysis and Results}

\subsection{PDS Data Analysis}

\begin{figure}[htb]
\psfig{figure=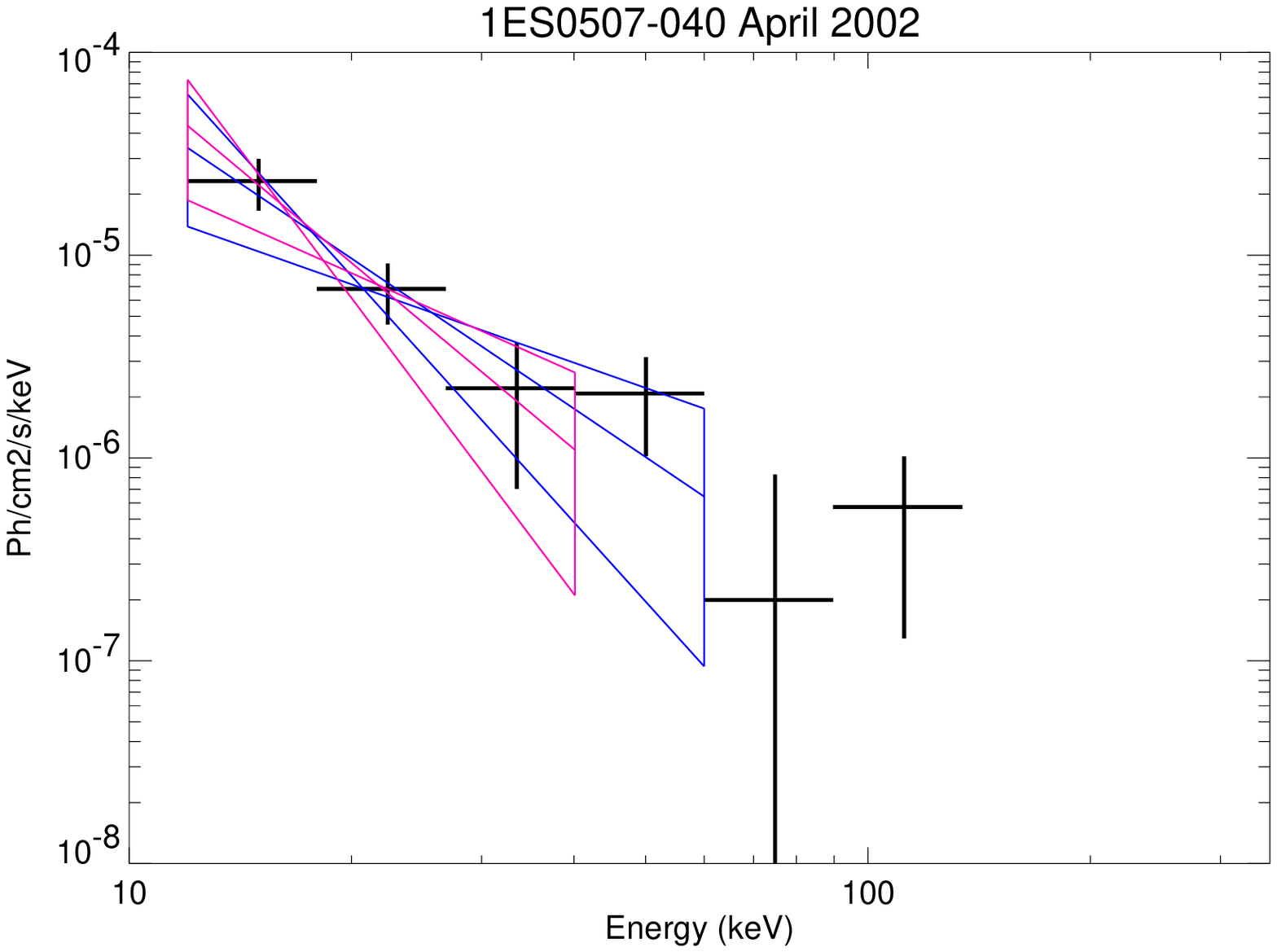,width=6.5cm}
\psfig{figure=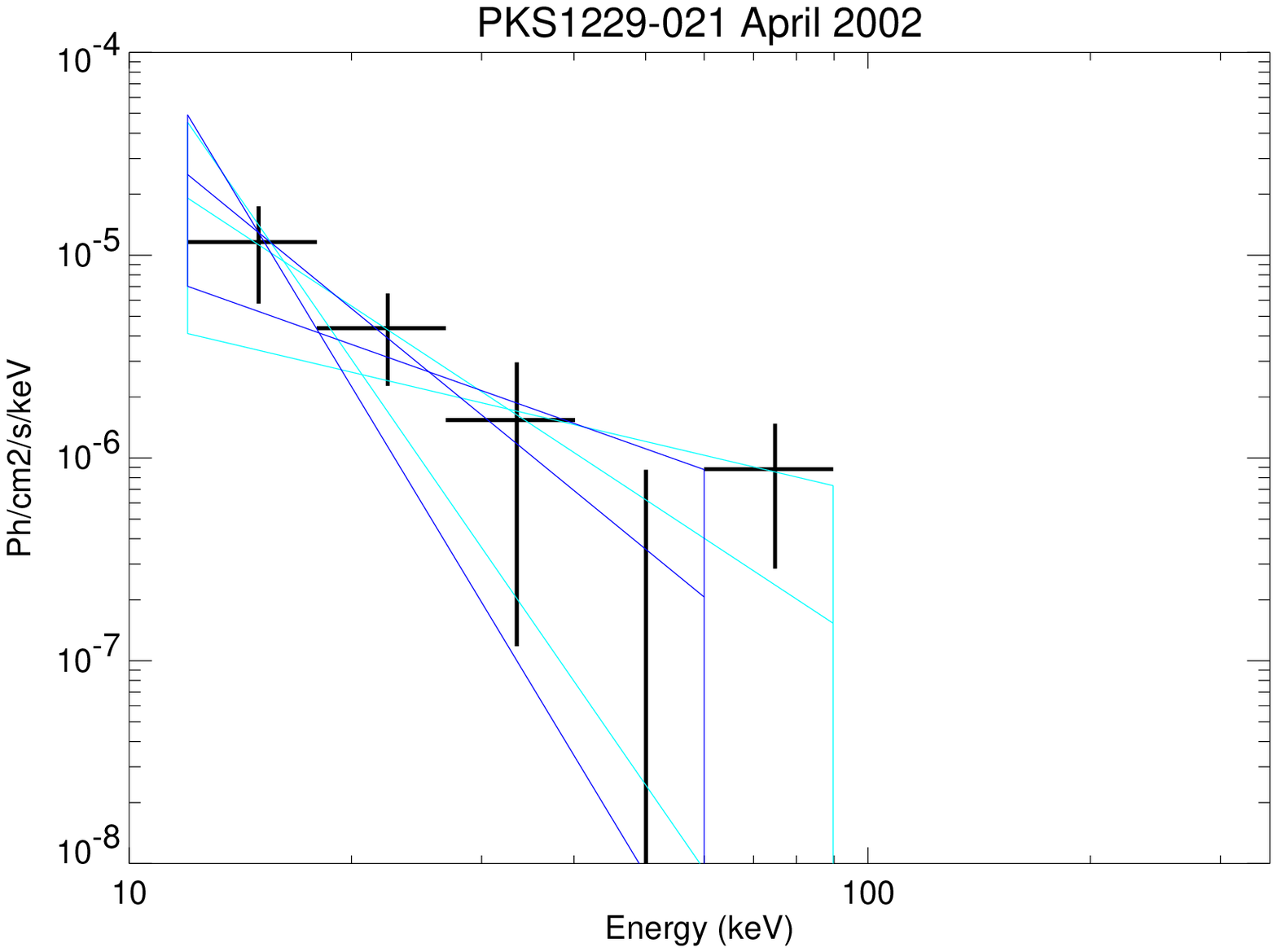,width=6.5cm}
\caption{The PDS spectra of ({\it Top}) 1ES 0507-040 April 2002 
and ({\it Bottom}) PKS1229-021 April 2002. The bow-ties indicate
the 90\% confidence correlated ranges of the slope and normalization
of power law fits using a few different bin ranges.}
\label{pds}
\end{figure}

We have performed the reduction of PDS data with XAS using standard
prescriptions (Chiappetti et al. 1999) inclusive of spike rejection, PSA
correction, unit-by-unit subtraction of the background spectra taken
with the collimator in offset position, and then summing the net
spectra of the 4 units with gain equalization and contextual
rebinning.
In order to derive points for a SED we rebinned the data in some
combinations of logarithmic bins and fitted the PDS data alone
(see Figure~\ref{pds}) with a power law, computing
the 90\% confidence intervals on the spectral index, and using the
correlated value of the normalization.

\subsection{Results}

{\bf 1ES0507-040}: this BL Lac object was discovered by EINSTEIN and was
the target of a previous {\it Beppo}SAX observation in a program aimed at
studying the spectral properties of a sample of X-ray selected BL Lacs
(Beckmann et al. 2002). It had been detected also in the ROSAT All-Sky
Survey (RASS).

\begin{figure}[htb]
\psfig{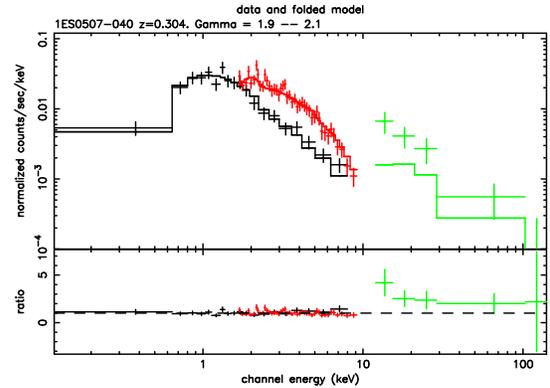} 
\caption{{\it Beppo}SAX spectrum of 1ES0507-040: LECS and MECS 
data (dark grey) are from the 11-12/02/1999 observation, (10 ks for LECS and PDS
and 20 ks for MECS) in which the PDS detection was marginal. The longer
2002 observation with the PDS (37 ks) is shown in light grey.}
\label{0507}
\end{figure}

Since the PDS count rate of the 2002 observation is consistent with that 
measured in 1999 ($0.047\pm 0.019$ vs $0.044\pm 0.020$ cts/s; see
Figure~\ref{pds}({\it Top}) for the April 2002 spectrum), it is
appropriate to fit both observations together, exploiting the larger
range observed in 1999 due to the LECS and MECS instruments. The best fit
spectral index for a power law in the range 0.1-50 keV with
$N_H=9.15\times 10^{20}$ cm$^{-2}$ (frozen at the Galactic
value) is $\Gamma = 2.03$ (1.95-2.13). The PDS normalization has been fixed
at the canonical value of 0.8$\times $MECS normalization. The PDS
spectral slope is consistent with that at lower energies at 90\%
confidence, while the flux is about a factor 3 higher. The global fit is
formally acceptable ($\chi ^2= 1.06$ with 76 d.o.f.) due to the low
statistical weight of the PDS points. However {\it the PDS flux is
clearly in excess of the fit prediction} (see Figure~\ref{0507}).

\begin{figure}[htb]
\centerline{\psfig{figure=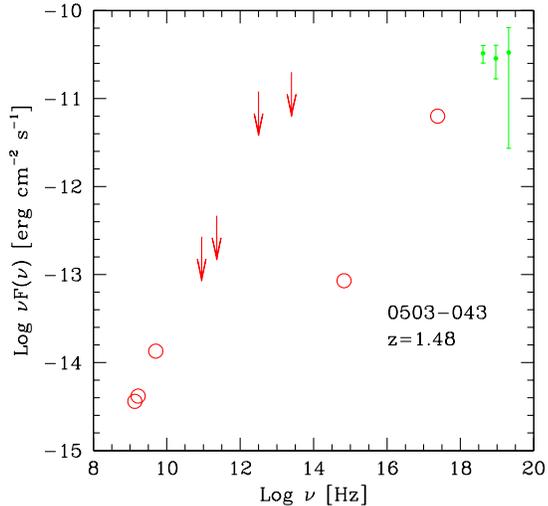,width=8.0cm}} 
\caption{Spectral Energy Distribution of S3 0503-04. The PDS flux has
been calculated with the appropriate off-axis correction (0.15),
assuming that the totality of the emission is produced by S3 0503-04.}
\label{0503}
\end{figure}

A plausible reason for this excess is that PDS data of 1ES0507-040
are contaminated by the emission from another source present in the large
PDS Field of View (about 1 deg). An inspection of catalogs shows that 
three possible contaminant sources are located within 1 deg from 1ES0507-040:

The radio source S3 0503-04, located at $\sim $1 deg from 1ES0507-040,
had been identified with a BL Lac of unknown redshift, and later shown to
have strong emission lines at $z=1.481$ (Veron 1994). The radio
morphology is compact (Stanghellini et al 1990). The radio spectrum is
flat, thus implying a blazar nature. In the X-rays the situation is less
clear: there is a 3$\sigma$ detection with HEAO A-2 (Della Ceca et
al. 1990) at fluxes $\sim 10^{-11}$erg cm$^{-2}$ s$^{-1}$, but this is not 
confirmed by the RASS data ($F_{0.2-2.4} <$ few
$10^{-13}$ erg cm$^{-2}$ s$^{-1}$ ). A possibility to account for the
large difference between the optical/soft X-ray luminosity and the hard
X-ray luminosity is to admit that low frequency radiation is heavily
absorbed. The possible SED is reported in Fig.~\ref{0503}. Further
X-ray observations are needed to investigate this interesting
possibility.

The other two bright {\it RASS} sources in the vicinity of 1ES0507-040 are
identified with G stars (HD 293857 at 25$^{\prime}$ and HD 32704 at 
70$^{\prime}$).  They
are unlikely to contribute to the PDS unless very peculiar (indeed HD
293857 is a lithium star, with $L_X\sim 10^{30}$ erg/s) or binaries.\\

\begin{figure}[htb]
\centerline{\psfig{figure=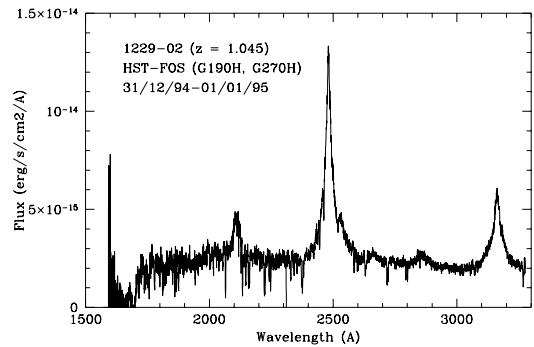,width=7.5cm}} 
\caption{HST FOS spectrum of PKS 1229-021 obtained by combining 2
spectra taken with grism G190H on 31 Dec 1994 with a grism G270H spectrum
taken on 01 Jan 1995. The fluxes are corrected for Galactic extinction
($E_{B-V} =0.032$).  Prominent broad
Ly${\alpha} $, Ly${\beta} $, CIV and weak SiIV emission are detected.}
\label{hst}
\end{figure}

\begin{figure}[htb]
\centerline{\psfig{figure=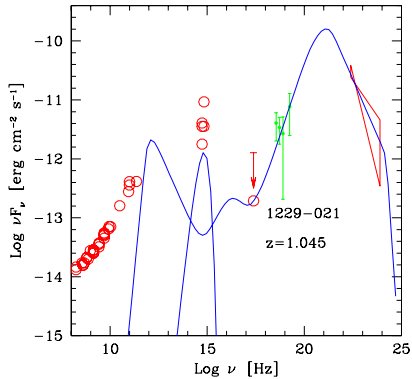,width=6.0cm}} 
\caption{Spectral Energy Distribution of PKS1229-021. Multiwavelenght
data are
taken from the literature, while PDS data are from this paper. The solid
line reports the synchrotron - Inverse Compton emission (calculated
according to the model described in Ghisellini, Celotti \& Costamante
2002) assuming the following parameters: $\delta =14.3$, $B=2.5$ G, $L_{\rm
inj}=9\times 10^{43}$ erg/s. The emitting region is a cylinder with
radius $R=2\times 10^{16}$ cm and thickness $1.5\times 10^{15}$ cm. The
disk radiation (bump at $10^{15}$ Hz, $L_d=1.2 \times 10^{46}$, 10
times the line luminosity) is reprocessed by the BLR clouds located at
$2 \times 10^{17}$ cm.}
\label{1229}
\end{figure}

{\bf PKS 1229-021}: this is a bright quasar (z=1.045), detected by EGRET in
the $\gamma$-ray and with a steep spectrum ($\Gamma =2.85$) (Hartman et
al. 1999). The optical
spectrum shows luminous emission lines (eg Fig.~\ref{hst}). 

We have retrieved from the HST FOS archive 2 spectra taken on Dec 31,
1994 with the G190H grating (1500-2300 \AA) and one taken on 01 Jan, 1995
with the G270H grating (2200-3300 \AA).  The spectrum, reported in
Fig.~\ref{hst}, has been corrected for the Galactic extinction (E$_{B-V}$ =
0.032, Schlegel et al. 1998).  Strong emission lines are detected,
corresponding to Ly$\alpha$, Ly$\beta$, Si IV, C IV. 
After removal
of the emission lines, a power-law fit of the combined spectra over the
1500-3300 \AA\ range yields an index $\alpha_\nu = 1.12 \pm 0.08$ ($f_\nu
\propto \nu^{-\alpha_\nu}$).

In the X-rays the source has been studied with ROSAT (RASS).  We have
constructed the overall SED, shown in Figure~\ref{1229}, using data
from the literature and our PDS data (see Figure~\ref{pds} {\it
Bottom}). We have reproduced the SEDs with the synchrotron-IC model of
Ghisellini, Celotti \& Costamante (2002).  We refer to that paper for a
full description of the model. The values of the basic physical
parameters are reported in the figure caption.

\section{Discussion and Conclusions}
The last {\it Beppo}SAX observations of blazars confirm the importance of
exploring the hard X-ray band in the study of blazars, as initiated by
{\it Beppo}SAX.

In PKS1229-021 the PDS spectrum allows us to constrain the position of
the high energy peak and to infer the basic physical parameters of the
relativistic jet.

The case of 1ES 0507-040 is more complex, due to a possible
contamination in the PDS spectrum by another source in the field of view. 
The
identification of the contaminant is difficult: based on the soft
X-ray emission the possible candidates are two active stars and a
peculiar radio-loud AGN. Future study with {\it INTEGRAL} and/or
new X-ray data are required to clarify the issue.

\end{document}